%Paper: hep-ph/9505442
%From: PINEDA@ecm.ub.es
%Date: Wed, 31 May 1995 20:45:26 +0000 (BCN)

\magnification=1200
\def\vsl{\rlap\slash{v}}
\overfullrule=0pt
\hfill UB-ECM-PF-95/6
\bigskip\bigskip
\centerline{\bf Matching the HQET}
\centerline{\bf to}
\centerline{\bf Coulomb-type bound states}
 \vskip 1.5cm
\centerline{A. Pineda}
\vskip 0.2cm
\centerline{and}
\vskip 0.2cm
\centerline{J. Soto}
\vskip 0.5cm
\centerline{Departament d'Estructura i Constituents de la Mat\`eria}
\centerline{and}
\centerline{Institut de F\'\i sica d'Altes Energies}
\centerline{Universitat de Barcelona}
\centerline{Diagonal 647, 08028 Barcelona, Catalonia, Spain }
\vskip 1cm

\centerline{\bf Abstract}
\bigskip   \bigskip

We show that the infinite imaginary anomalous dimensions of certain
heavy quark-antiquark currents in the HQET arise due to an inappropriate
commutation of two limits, namely coinciding velocities and infinite
cut-off. This commutation is not apparent when dimensional
regularization
is used, but it can be made manifest in a cut-off regularization.

We argue that operators containing heavy quark and antiquark
fields with the same
velocity in the HQET must not be matched to perturbative QCD but to
Coulomb-type
bound state contributions. We show how to do so at one loop level.

\vfill
\footnote{}{e-mail: pineda at ecm.ub.es, soto at ecm.ub.es }

 \eject

\def \vp{v^{\prime}}
\def \kp{k^{\prime}}

{\bf 1. Introduction}

\bigskip
The so called Heavy Quark Effective Theory (HQET) has become a standard
tool to study hadrons composed of a single heavy quark [1] (see [2] for
reviews).
 For these hadrons
there are only two relevant scales, $m_{Q}$ the mass of the heavy quark
and $\Lambda_{QCD}$. If $m_{Q}$ is large enough one can use perturbative
QCD to calculate contributions coming from momenta $p$,
$p>>\mu >>\Lambda_{QCD}$ where $\mu $ is a cut-off such that $m_{Q}
>>\mu $. For momenta
$p << \mu $ one may not carry out reliable perturbative calculations but
one can use HQET, the symmetries of which relate
different form factors and matrix elements. Since there is no
relevant physical
scale between $m_{Q}$ and $\Lambda_{QCD}$, one can match HQET to QCD
at the scale $m_{Q}$ order by
order in perturbation theory. The Wilson coefficients thus
obtained may be run to lower scales
 by using standard renormalization
group techniques. Roughly speaking, for the above mentioned hadrons a
current in
full QCD can be written as a current in the HQET multiplied by a Wilson
coefficient which depends on $\alpha (m_{Q})$, $\alpha (\mu)$ and
an anomalous dimension which is computable perturbatively.
\bigskip
It was initially thought that the HQET had no applications to hadrons
containing more than one heavy quark, in particular to heavy quarkonium.
The reason is that those hadrons can be understood as weak coupling
Coulomb Type Bound States (CTBS), which cannot be obtained from the
static
approximation.
Unlike in the heavy-light systems, the short distance effects are
dominant in heavy quarkonium. Long distance (non-perturbative)
effects have been traditionally simulated by the inclusion of a smoothly
varying background  field which eventually gives
rise to contributions depending on the gluon condensate [3,4]  (see
[5] for a recent work).
Although this method encodes important features of the
non-trivial QCD vacuum, it is not obvious that it exhausts all possible
non-perturbative contributions\footnote{$^{\ast}$}{That would
be the case if some non-perturbative contributions could not be
expressed
in terms of local condensates. This situation has been recently
discussed in [6].}.

\bigskip

 Surprisingly enough, the HQET seems to know about quarkonium.
On the one hand, the infinite imaginary anomalous dimensions found in
operators with a heavy quark and a heavy antiquark when the velocities
coincide have been related to the static interquark potential [7]. On
the other hand, quarkonia states appear to be related to Goldstone modes
arising from the spontaneous breaking of $U(4N_{hf})\rightarrow
U(2N_{hf})\otimes
U(2N_{hf}) $ in a HQET containing quarks and antiquarks with the same
velocity [8,9].
 Recently, it has been shown that HQET techniques can indeed
be used to parametrize certain non-perturbative contributions in heavy
quarkonium [9]. These contributions correspond to quantum fluctuations
that put quark and anti-quark in quarkonium almost on shell when
the relative three momentum is much smaller than
the inverse Bohr radius.
 In this kinematical situation the quark and the antiquark move almost
with the same three velocily, namely the quarkonium three velocity.
Non-perturbative contributions in this
region are important and can be described using HQET.
However, in this situation the HQET must not be matched to QCD order by
order in perturbation theory. One should
keep in mind that
 in addition to $m_{Q}$ and $\Lambda_{QCD}$
 a relevant scale,
namely the Bohr radius $a_{n}\sim n/(\alpha m_{Q})$, laying
between $m_{Q}$ and $\Lambda_{QCD}$ enters the problem.
The HQET here must be matched to
CTBS contributions, which
arise from an infinite summation of one gluon exchange diagrams of
QCD, the so-called ladder diagrams. The matching above  has
been carried out at tree level in [9].
 \bigskip
The main goal of this note is to explain how such a matching can be
carried
out at one loop order. The fact that HQET must be matched to CTBS
contributions and not to perturbative QCD is related to the
infinite
imaginary anomalous dimensions found in [7,10]. We shall also see that
those
  are an
artifact
due to an inappropriate commutation of two limits,
namely coinciding
velocities and infinite cut-off.
\bigskip
In order to avoid confusion, we would like to properly locate our work
with respect to the standard approach by
 effective theory techniques to heavy
quarkonium, namely the so called Non-Relativistic QCD (NRQCD) [11].
NRQCD is an
 effective theory for momenta of the order of the inverse Bohr radius or
smaller, in
contradistintion to HQET which is an effective theory for momenta of
the
order of $\Lambda_{QCD}$. NRQCD also admits an expansion in $1/m_{Q}$
but momentum counts as inverse Bohr radius $m_{Q}\alpha$ whereas energy
as $m_{Q} \alpha^2$. There is no flavour symmetry although spin
symmetry
remains. Contributions coming from momentum larger than the inverse Bohr
radius can be calculated perturbatively in $\alpha (m_{Q})$. They give
rise to Wilson coefficients when matching NRQCD to QCD.
Momenta
of the order of the inverse Bohr radius give rise to weak coupling
CTBS. However, there are also momenta of the order
of $\Lambda_{QCD}$ which bring in non-perturbative contributions.
One may
try to evaluate both perturbative and non-perturbative
contributions by lattice simulations starting from NRQCD [12]. We
believe,
however, that one can proceed analytically one step further in the
separation of
perturbative and non-perturbative contributions, at least for very heavy
quarks. We have been able to
 describe some of these non-perturbative effects by using
HQET.
Here, we would like to present in somewhat more detail how the matching
of HQET to CTBS contributions can be carried out. It should be clear
that this is
quite different from describing quarkonia states starting from the HQET
(static limit) which is known to be a hopeless task.
\bigskip
We organize the paper as follows. Section 2 contains a technical
discussion on the integral from which the infinite imaginary anomalous
dimensions arise. In Section 3 we sketch how the HQET can be matched to
CTBS.
The result depends crucially on the correct evaluation of integrals
analogous to the one
analysed in Section 2. In Section 4 we carry out the matching at one
loop within a proper setting.
Section 5 is devoted to the
conclusions.

\bigskip\bigskip
{\bf 2. Dimensional regularization versus three-momentum cut-off
regularization}
\bigskip
Consider the following integral, which arises in the calculation of
anomalous dimensions of heavy-heavy currents in the HQET.
 $$
I_{\pm}(v.\vp, v.k,\vp.\kp)
:= -i\int {d^4q\over (2\pi)^4}
 {1\over
v.(q+k)+i\epsilon}
 {1\over
\pm\vp.(q+\kp)+i\epsilon}
 {1\over
q^2+i\epsilon}.  \eqno (2.1)
 $$
$$v^0\;,\;{v^0}^{\prime}\; > 0\,.$$
$I_{\pm}$ above is ultraviolet divergent. If we use dimensional
regularization,
 i.e.
 $I_{\pm} \rightarrow I_{\pm}^{D}$,
 we obtain
$$\eqalignno{
  I_{+}^{D}(v.\vp, v.k,\vp.\kp) &=
 {-1\over 4-D}
{1\over 4 \pi^2}g(\vp .v)
+{\rm finite \; terms},
&(2.2) \cr
  I_{-}^{D}(v.\vp, v.k,\vp.\kp) &=
 {1\over 4-D}
{1\over 4 \pi^2}(g(\vp .v) -i{\pi\over \xi})
+{\rm finite \; terms}
        \quad\quad v\not=\vp, & (2.3)
\cr
  I_{-}^{D}(1, v.k,v.\kp) &=
{1\over 4 \pi^2}[
 {1\over 4-D} + {1\over v.(k-\kp) +i\epsilon}(
v.k\ln{\mu\over \vert v.k\vert}
-v.\kp\ln{\mu\over \vert v.\kp\vert}
\cr & +v.k\theta (v.k)i\pi
-v.\kp\theta (-v.\kp)i\pi )],
      & (2.4)    \cr
   }
$$
where
$$
\xi:=\sqrt{(\vp .v)^2-1}\,, \quad\quad
g(\vp .v):={1\over 2\xi}\ln{
\vp .v +\xi      \over
\vp .v -\xi }. \eqno (2.5)
$$
The anomalous dimensions of certain currents may be obtained from
the coefficients of the divergent terms in (2.2)-(2.4).
We have calculated (2.2)-(2.4) directly in
Minkowski space by carrying
out first the integral over the zero momentum component. No Feynman
parametrization or similar tricks which may lead to ill-defined
expressions have been used. Keeping track of the $i\epsilon$
is crucial to avoid ambiguities.
Notice that $$
  \lim_{\vp\rightarrow v} I_{-}^{D}(v.\vp, v.k,\vp.\kp) \not=
  I_{-}^{D}(1, v.k,v.\kp) \,,
\eqno (2.6)
$$
which is going to be important in the following.
The lhs becomes infinite imaginary when $v\rightarrow v^{\prime} $ (i.
e. $\xi\rightarrow 0$ and $g(\vp .v) \rightarrow g(1)=1$). However, the
rhs gives a finite
real anomalous dimension.
The original way of obtaining
$I_{-}^{D}(v.\vp, v.k,\vp.\kp)$ [7,10] was through the analytic
continuation of
$I_{+}^{D}(v.\vp, v.k,\vp.\kp)$  to negative ${v^0}^{\prime}$, more
precisely
 $v.\vp\rightarrow -v.\vp -i\epsilon $
, which coincides with
the result (2.3). This procedure requires $v\not=\vp $ and hence the
discontinuous limit (2.6) cannot be uncovered
\footnote{$^\ast$}{Note
that the analytic continuation
 $v.\vp\rightarrow -v.\vp -i\epsilon $
transforms one HQET
(two heavy quark fields with velocities $v$ and $\vp$
respectively)
 into a different HQET
(a heavy quark fields and a heavy antiquark field with velocities $v$
and $\vp$ respectively). This is quite different from the usual analytic
continuation which relates the s and t channel within a given
relativistic QFT.
In fact, it only makes sense if we regard both HQETs as emanating from a
single QFT, namely QCD, where crossing holds.}.

\bigskip

 It is difficult to believe that physics is
discontinuous when $\vp$ is close to $v$ so the question arises which
expression (2.3) or (2.4), if any of them, is relevant in that
situation.
In order to gain a better understanding of the non-commuting limit
(2.6), let us
consider a spherically symmetric three-momentum cut-off regularization,
 i.e.
 $I_{\pm} \rightarrow I_{\pm}^{\mu}$ where
$ \int d^4q \rightarrow \int d(v.q) \int^\mu d(e^i.q) \quad(e^i.v =
0)\,.$  We obtain

$$I^{\mu}_{+}(\vp .v,v.k, \vp .\kp ) =
-{1\over 4 \pi^2}g(\vp .v)
\ln{\mu}
        +{\rm finite\; terms}
\eqno (2.7) $$
and hence the anomalous dimension agrees with (2.2).
The ultraviolet behaviour of $I^{\mu}_{-}$ turs out to be qualitatively
different in the following two situations
$$ \eqalign{
 i)
\quad\quad\quad\quad\quad
& {\xi\mu \over v.\vp} >> \vert v.k-{\vp.\kp \over v.\vp}
\vert \,,
\cr
 ii)
\quad\quad\quad\quad\quad
& {\xi\mu \over v.\vp} << \vert v.k-{\vp.\kp \over v.\vp}
\vert \,.\cr }  \eqno (2.8)
 $$
This can be seen from the expression
$$\eqalign{
&I^{\mu}_{-}(\vp .v,v.k, \vp .\kp ) =
-{1\over 8 \pi^2}{1\over\xi}\int^{\mu}_{0} dq \Biggl\{ \left[
{1\over q+v.k+i\epsilon}+
{1\over q-v.k-i\epsilon} \right] \cr & \times
\ln{
{-{\xi q \over v.\vp}+ v.k-{\vp.\kp \over v.\vp}+i\epsilon\over
{\xi q \over v.\vp}+ v.k-{\vp.\kp \over v.\vp}+i\epsilon}}
-{1\over q+v.k+i\epsilon}
\ln{
{{\xi q \over v.\vp}+q+{\vp.\kp \over v.\vp}-i\epsilon\over
-{\xi q \over v.\vp}+q+{\vp.\kp \over v.\vp}-i\epsilon}}\Biggr\}\,.}
\eqno (2.9)
$$
Recall that always $\vp.\kp, v.k << \mu$. In the case i) we obtain
$$I^{\mu}_{-}(\vp .v,v.k, \vp .\kp ) =
\ln{\mu}
{1\over 4 \pi^2}(g(\vp .v) -i{\pi\over \xi})
        +{\rm finite\; terms,}
\eqno (2.10) $$
which leads to the same result as (2.3) for the anomalous dimension. In
the case ii) we obtain
$$ \eqalign{
I^{\mu}_{-}(\vp .v,v.k, \vp .\kp ) =&
{1\over 4 \pi^2}
  {1\over v.k-\vp.\kp +i\epsilon}(2\mu +
v.k\ln{\mu\over \vert v.k\vert}
-\vp.\kp\ln{\mu\over \vert \vp.\kp\vert} \cr &
+v.k\theta (v.k)i\pi
-\vp.\kp\theta (-\vp.\kp)i\pi ).              }
\eqno (2.11)
$$
When $v\rightarrow v^{\prime}$, $\xi\rightarrow 0$ we are
always in the case ii) for any finite cut-off $\mu$. From the
logarithmically divergent terms in (2.11) we identify the same
anomalous dimension as in (2.4), but there is an additional
 linearly divergent term. This
 term cannot appear in (2.4)
 due to the dimensional regularization prescription of putting scale
invariant integrals to zero.
{}From (2.8) and (2.11) it is apparent that the relevant result for
$I_{-}$ when the velocities of the quark
and antiquark are close has to do with (2.4) and not with the limit
$\vp\rightarrow v$ of (2.3).
The point is that a physical scale $\mu$ must definitely be
introduced to separate the momenta from which we can start using HQET.
With $v\not=v^{\prime}$ there are two scales, namely $\mu$ and $\mu
(v.v^{\prime}-1 )$. Both scales are large for $\mu$ large as far as
 $v\not=v^{\prime}$. In that case dimensional regularisation and the
three-momentum cut-off regularisation are equaly convenient and they
give the same
result for the anomalous dimension (which may be imaginary).
However, when $\vp\rightarrow v$ the second scale goes from being large
to vanishing. Since
 dimensional
regularization is only able to provide a single scale (the subtraction
point), the only way it has to show us
 the brutal change in the second scale is by means of the
discontinuous limit (2.6). The three-momentum cut-off is more
convenient in this case. It allows us to examine the relevant limit
in terms of physical scales (2.8) and to uncover the qualitative
difference in the UV behaviour of $I_{-}$ for $\vp$ much different
from $v$ and for $\vp$ similar to $v$, that is the linear divergence in
(2.11).

\bigskip

 The linear divergent term in (2.11) has indeed
far reaching consequences: it is non-local.
The non-local cut-off dependence indicates that the HQET for quarks and
antiquarks with
the same velocity does not behave as a renormalizable quantum
field theory, when we calculate Green functions involving
currents with quark and antiquark fields
\footnote{$^{\dagger}$}{This is not in
contradiction with ref. [13]
where it was found that HQET, including quarks and antiquarks with the
same velocity, is renormalizable
since the renormalization of currents was not considered.}.
In this situation the HQET would be of little use
if we did not know the
high momentum physics which lies beyond the cut-off. Fortunately,
we do
know the high momentum physics which lies beyond the cut-off:
Coulomb-type
bound states. It is not obvious, however, whether one can actually match
these contributions in the HQET with
 contributions
arising from CTBS.
 We devote the remaining sections to this goal.

\bigskip

After having found a non-local linear divergence in (2.11) we may wonder
at which extend the result that we obtain depends on the
regularization that we use. For $v=v^{\prime}$ and taking the Coulomb
gauge in (2.1) (i.e. $1/(q^2+i\epsilon ) \rightarrow -1/(e^{i}.q)^2 $)
the following statements are easy to check. For both
$U(N_{hf})$ and gauge invariant regulators of the kind
$$
{1\over \vsl v.D +i\epsilon}
\rightarrow
{1\over \vsl v.D +i\epsilon}f(
-{D_{i}^2\over \mu^2})
\eqno (2.12)
$$
where $f(x)\rightarrow 0$ strong enough when $x\rightarrow \infty $ and
$f(0)=1$, the non-local linear divergence and the pole in (2.11) remain.
Regulators of the kind
$$
{1\over \vsl v.D +i\epsilon}
\rightarrow
{1\over \vsl v.D -\mu (
-{D_{i}^2\over \mu^2})^n
+i\epsilon}
\eqno (2.13)
$$
which do not respect the $U(4N_{hf})$ symmetry, give different
branching
point singularities depending on $n$ instead of a simple pole. $
U(4N_{hf})$-invariant 'regulators' of the kind
$$
{1\over \vsl v.D +i\epsilon}
\rightarrow
{1\over \vsl\left( v.D -\mu (
-{D_{i}^2\over \mu^2})^n\right)
+i\epsilon}
\eqno (2.14)
$$
do not regulate the integral. Neither a Pauli-Villars regularization for
the heavy quark (antiquark) propagators does. Consequently the non-local
linear divergence in (2.11) appears to follow from the use of a
$U(4N_{hf})$-invariant regulator.
Indeed, the three dimensional cut-off
regularization respects the
$U(4N_{hf})$ symmetry.
We have checked that the Ward
identities for the heavy quark-antiquark current are fulfilled at one
loop level with no need to introduce any
counterterm\footnote{$^{\ast}$}{The three dimensional cut-off
regularization, however, breaks local $SU(3)$ gauge invariance
(Slavnov-Taylor
identities) which, nevertheless, can be restored by adding suitable
(non-gauge invariant) local counterterms. We have checked this stament
at one loop order.}.

                         \bigskip

To summarize,
for $\vp$ very different from $v$
(in the precise sense
(2.8))
we find the same imaginary anomalous dimensions as previously obtained
[7,10].
For $\vp$ close to $v$
(in the precise sense
(2.8)),
 we conclude that there is
no infinite imaginary anomalous dimension as claimed in [7], but a
non-local linear cut-off dependence. This non local dependence
cancels against
contributions coming from high energies (CTBS) once the matching has
been performed properly, as we shall see in Sect. 4.

                       \bigskip\bigskip
{\bf 3. Matching CTBS with HQET:
 getting the flavour.}
     \bigskip

Consider the following Green function to one loop in the HQET.
$$  G_{\Gamma}(k_1,k_2)=
\int d^4x_1d^4x_2 e^{-ik_1x_1-ik_2x_2}\langle 0
\vert T\left\{ \bar
h^{a} \Gamma h^{b}(0){\bar h}^{b\;i_1}_{+\alpha_1}(x_1)
h^{a\;i_2}_{-\alpha_2}(x_2)\right\} \vert 0 \rangle\,,\eqno (3.1)
 $$
$a,b...=1,...N_{hf}$ are flavour indices (repeated flavour indices
are not summed up).
Colour indeces are not explicitly displayed
in the colour singlet currents. Otherwise they will be denoted by $i_1$,
 $i_2$, ...$=1 .... N_{c}$,
 $N_{c}$ being the number of colours. $\alpha_1$, $\alpha_2$ are spinor
indices. $h_{+}^{a}$ annihilates quarks whereas $h_{-}^{a}$ creates
antiquarks. $h_{\pm}^{a}$ have two independent components each. We
combine them often into a four independent component field
$h^{a}:=h_{+}^{a}+h_{-}^{a}$ ($h_{\pm}= p_{\pm}h; p_{\pm}= (1 \pm
\rlap\slash{v})/2$).
 Let us choose the Coulomb
gauge ($\partial_{i}B^{i}=0$, $x^{i}:=e^{i}.x$, $B^{i}:=e^{i}.B$,
$e^{i}.v=0$, $i=1,2,3$). We obtain at one loop level (see Fig. 1)
 $$\eqalignno{
  G_{\Gamma}(k_1,k_2)=&
  G_{\Gamma}^0(k_1,k_2)+
  G_{\Gamma}^{11}(k_1,k_2)+
  G_{\Gamma}^{12}(k_1,k_2)+
  G_{\Gamma}^{13}(k_1,k_2).
& (3.2) \cr
  G_{\Gamma}^0(k_1,k_2)=&
 (p_{-}\Gamma p_{+})_{\alpha_2\alpha_1}
\delta_{i_1 i_2}
{1\over v.k_2 +i\epsilon} \,
{1\over v.k_1 +i\epsilon} \,.  & (3.3) \cr
  G_{\Gamma}^{11}(k_1,k_2)=&
  {C_{f}\alpha\over \pi}\mu
 (p_{-}\Gamma p_{+})_{\alpha_2\alpha_1}
\delta_{i_1 i_2}
({1\over v.k_2 +i\epsilon})^2 \,
{1\over v.k_1 +i\epsilon} \,.  &(3.4) \cr
  G_{\Gamma}^{12}(k_1,k_2)=&   {C_{f}\alpha\over \pi}\mu
 (p_{-}\Gamma p_{+})_{\alpha_2\alpha_1}
\delta_{i_1 i_2}
({1\over v.k_1 +i\epsilon})^2 \,
{1\over v.k_2 +i\epsilon} \,.  &(3.5) \cr
  G_{\Gamma}^{13}(k_1,k_2)=&
 - {C_{f}\alpha\over \pi}2\mu
 (p_{-}\Gamma p_{+})_{\alpha_2\alpha_1}
\delta_{i_1 i_2}
{1\over v.k_1 +i\epsilon} \,
{1\over v.k_2 +i\epsilon} \,
{1\over v.(k_1+k_2) +i\epsilon} \,.
 &(3.6) \cr                     }
$$
$$ C_{f}:= {N_{c}^2-1\over 2N_{c}}\,.$$
The linear divergences in (3.4) and (3.5) are local and may be
understood as mass renormalizations. The linear divergence in (3.6) is
non-local and appears from an integral analogous to the one discussed
in the previous section.
If we take the Lorentz-Feynman gauge additional logarithmic dependences
in $\mu$ arise. The choice of the Coulomb gauge is not only a matter of
convenience. As we have already mentioned we should match the HQET
results to CTBS contributions. The latter are
always calculated in the Coulomb gauge. Then, in
order to match with the results obtained by the conventional treatment
of CTBS, using the Coulomb gauge in the HQET is mandatory.

\bigskip

In ref. [9] it was shown that, at leading order in $\mu a_{ab,n}$,
$a_{ab,n}$ being the Bohr radius, $
G_{\Gamma}^0(k_1^{\prime},k_2^{\prime}) $ can be matched with

 $$  G_{\Gamma}(p_1,p_2):=\int
d^4x_1d^4x_2 e^{ip_1x_1+ip_2x_2} \langle 0\vert T\left\{ {\bar
Q^{a}} \Gamma Q^{b}(0)
{\bar Q}^{b i_1}_{\alpha_1}(x_1)
Q^{a i_2}_{\alpha_2}(x_2)
 \right\} \vert 0 \rangle \,. \eqno (3.7) $$
$$ p_1=-m_{b}v+{m_{b}\over m_{ab}} E_{ab,n} v-k_1^{\prime} \quad , \quad
p_2=-m_{a}v+{m_{a}\over m_{ab}} E_{ab,n} v-k_2^{\prime} \,,\quad
 $$
$$
m_{ab}:=m_{a}+m_{b}  \quad\quad , \quad\quad
 k_1^{\prime}\; ,\; k_2^{\prime} \rightarrow 0 \,,
$$
under the assumption that CTBS dominate the
above Green function. The precise relation reads
 $$  G_{\Gamma}(p_1,p_2) =C_{\Gamma}
  G_{\Gamma}^0(\kp_1,\kp_2) \quad\quad ,\quad\quad
  C_{\Gamma}=
\tilde \Psi_{ab,n}^{\ast}(0)
 \Psi_{ab,n}(0).                      \eqno (3.8)
$$
$E_{ab,n}$, $\Psi_{ab,n}(\vec x)$ and $\tilde \Psi_{ab,n}(\vec k)$ are
the energy, the coordinate space wave function and the momentum space
wave function
respectively of a Coulomb-type state with principal quantum number $n$.
We call (3.8) tree level matching, meaning that the CTBS contribution
matches with a tree level diagram of the HQET.

\bigskip

 Let
us concentrate on
$  G_{\Gamma}^{13}(k_1,k_2) $. Anologously to (3.8), we would like to
find
a contribution coming from CTBS which matches with this contribution in
the HQET.
In order to do so, one way or another, we have to single out a gluon
propagator from
the infinite set in the ladder diagrams.
 At the risk of having some double counting, which we
shall overcome in next section, we may proceed as follows. We start as
if we computed (3.7) at one loop in QCD (or NRQCD), namely
$$
 \eqalign{
  G_{\Gamma}^{13}(p_1,p_2)=&\int d^4x_1d^4x_2 e^{ip_1x_1+ip_2x_2}
\langle 0\vert T\{ {\bar
Q^{a}} \Gamma Q^{b}(0)
 \cr & \times
\int d^4zd^4 z^{\prime} i^2
(\bar Q^{b}\gamma_{\mu}B^{\mu}  Q^{b})(z)
(\bar Q^{a}\gamma_{\nu}B^{\nu}  Q^{a})(z^{\prime})
 \cr & \times
{\bar Q}^{b i_1}_{\alpha_1}(x_1)
Q^{a i_2}_{\alpha_2}(x_2)
 \} \vert 0 \rangle \,. } \eqno (3.9)
$$
$p_1$ and $p_2$ are as in (3.7). At this stage we introduce resolutions
of the identity approximated
by CTBS between the currents and the interaction terms. Notice that
since we use the Coulomb gauge the two
interaction currents become a single time four quark operator. We then
have (see Fig. 2)
 $$
 \eqalignno{
  G_{\Gamma}^{13}(p_1,p_2)=&\int d^4x_1d^4x_2 e^{ip_1x_1+ip_2x_2}
\int d^4z d^4z^{\prime} i^2
\theta (-max(z^0,{z^{\prime}}^0)) &\cr & \times
\theta (min(z^0,{z^{\prime}}^0)-
max(x_1^0,x_2^0) )
\sum_{n,m}
\int {d^3\vec P_{n} \over (2\pi )^3 2P_{n}^0}
\int {d^3\vec P_{m} \over (2\pi )^3 2P_{m}^0}
\langle 0\vert
 {\bar
Q^{a}} \Gamma Q^{b}(0)
\vert n\rangle
 &\cr & \times
\langle n\vert T\{(
\bar Q^{b}T^{r}\gamma_{\mu}  Q^{b})(z)(
\bar Q^{a}T^{r}\gamma^{\mu}  Q^{a})(z^{\prime})\}
\vert m\rangle
\langle m\vert T\{
{\bar Q}^{b i_1}_{\alpha_1}(x_1)
Q^{a i_2}_{\alpha_2}(x_2)
 \} \vert 0 \rangle \, &\cr& \times
g^2 \int {dq^0\over 2\pi}\int^{\mu} {d^3 \vec q\over (2\pi)^3}{i\over
{\vec q}^2}e^{-iq.(z-z^{\prime})}\,.
 & (3.10) \cr  }
$$
We always calculate in the non-relativistic limit
$a_{ab,n}m_{a}$, $a_{ab,n}m_{b}>>1$. In this limit
$\gamma_{\mu}\sim v_{\mu}$, which we have already assumed when we wrote
down the static component of the gluon propagator only.
If we constrain the gluon momentum to be smaller than $\mu$,
 we obtain at leading order in
 $\mu a_{ab,n}$
 ($\mu a_{ab,n} <<1$)
 $$
 \eqalign{
  G_{\Gamma}^{13}(p_1,p_2)= &
-\tilde \Psi_{ab,n}^{\ast}(0)
 \Psi_{ab,n}(0)
{C_{f}\alpha\over\pi}2\mu
 (p_{-}\Gamma p_{+})_{\alpha_2\alpha_1}    \cr & \times
\delta_{i_1 i_2}
{1\over v.\kp_1 +i\epsilon} \,
{1\over v.\kp_2 +i\epsilon} \,
{1\over v.(\kp_1+\kp_2) +i\epsilon} \,,        }
\eqno (3.11) $$
which matches
$  G_{\Gamma}^{13}(\kp_1,\kp_2) $
 in (3.6)
with the same Wilson coefficient $C_{\Gamma}$ as in (3.8). This is very
illustrative.
It tells us that the HQET at one loop reproduces the exchange of a
singled out soft gluon with momentum lower than $\mu$ in a CTBS. If we
add to (3.10) the momenta larger than $\mu$ for the singled out gluon
the result, of course, is $\mu$ independent. Thus we have identified the
large momentum contributions which cancel out the worrysome non-local
linear divergence in
$  G_{\Gamma}^{13}(k_1,k_2) $.

\bigskip

Although the exposition above contains the
essence of the matching between CTBS and HQET,
the procedure has to be refined. There are indeed two outstanding
problems.
First of all,             $
  G_{\Gamma}^{11}(k_1,k_2) $ and $
  G_{\Gamma}^{12}(k_1,k_2)        $
cannot be matched with (3.7).
 These
contributions correspond to one particle reducible diagrams and amount
to mass renormalizations of the
heavy quark
and the heavy antiquark respectively and independently. Contributions
like these do not
arise from a calculation of the kind (3.7)-(3.11), that is, from the
diagrams
$  G_{\Gamma}^{11}(p_1,p_2) $ and
$  G_{\Gamma}^{12}(p_1,p_2) $  in Fig. 2.
 The reason is that inside quarkonia quark
and antiquark are always communicating through the exchange
of gluons and hence one cannot get independent contributions for either.
 We can get ride of this missmatch by
 restricting our analysis to
local operators of the heavy quark and antiquark fields, as we shall do
in the next section.
 This is not an important restriction since
local operators are the objects which
actually arise in the calculations of physical observables.
 Second, and more important, there may be a problem of double counting.
On the one hand, we sum up an infinite
set of gluon propagators of arbitrary momentum in ladders to obtain the
Coulomb potential. On the other hand,  we single out a gluon
propagator, which we force to have small momentum, and treat it
perturbatively. (Is this gluon propagator
not already counted in the infinite set?). We show how to avoid this
problem in the next section.

 \bigskip\bigskip

{\bf 4. Matching CTBS with HQET: a proper setting}

\bigskip
Consider the Green function analogous to (3.1) but with quark and
antiquark fields at the same point in a current.
$$  G_{\Gamma\Gamma^{\prime}}(p):=
\int d^4x
e^{ipx}\langle 0\vert T\left\{
\bar Q^{a} \Gamma Q^{b}(0)
\bar Q^{b}\Gamma^{\prime} Q^{a}(x)
\right\} \vert 0 \rangle \,,\eqno (4.1)
$$
$$p=-m_{ab,n}v-k \quad ,\quad k\rightarrow 0 \,. $$

We introduce two cut-offs $\mu^{\prime}$ and $\mu$, both of them
 much
smaller than the inverse Bohr radius.
The first cut-off
$\mu^{\prime}$ separates two different kinematical situations, small
(high) relative three momentum, in the heavy quarks currents.
We denote them as on-shell (off-shell)
currents.
Our goal is reproducing
(4.1) in the HQET when one of the currents is almost on-shell.
$$  G_{\Gamma\Gamma^{\prime}}^{on}(p):=
\int d^4x
e^{ipx}\langle 0\vert T\left\{
(\bar Q^{a} \Gamma Q^{b})_{off}(0)
(\bar Q^{b}\Gamma^{\prime} Q^{a})_{on}(x)
\right\} \vert 0 \rangle \,,\eqno (4.2)
$$
The on-shellness will be implemented by cutting off the relative momenta
larger than $\mu^{\prime}$ in the wave function.
 The second cut-off $\mu$
separates
the gluon momenta between large and small.
 We assume
that the relevant
contributions from large gluon momenta are the infinite sums of one
gluon exchange diagrams (ladder diagrams) which give rise to
the CTBS.
For small gluon momenta we use perturbation theory
(no infinite summation)\footnote{$^{\dagger}$}{For $m_{Q}\rightarrow
\infty $ we can always assume that perturbation theory is applicable at
scales even lower than $m_{Q}\alpha$ since this quantity also goes to
infinity and $\alpha(m_{Q}\alpha)$ may still be small. This
is not so for the actual values of the $m_{b}$ and $m_{c}$.
Nonetheless once we find the structure of the matching perturbatively we
believe we may safely assume that the same structure holds when one
takes into account non-perturbative contributions as well.}.
Again we work at leading order in $\mu a_{ab,n}$ and in the
non-relativistic limit $m_{a}a_{ab,n}$, $m_{b}a_{ab,n} >> 1$.
Notice that double counting is explicitely avoided by the
introduction of the cut-off. This is a proper framework in which the
matching between CTBS and HQET can be carried out.
 A given order of perturbation theory for small momentum gluons
calculated this way corresponds
to the same order of perturbation theory in the HQET. We carry out the
matching at one loop below.

\bigskip

At zeroth order, i.e. no soft gluon propagator, we have

$$
G_{\Gamma\Gamma^{\prime}}^{on,0}(p)=
-i
N_{c}tr(p_{-} \Gamma p_{+} \Gamma^{\prime})
{\tilde \Psi^{\ast\,\,\mu}_{ab,n}}(0){\Psi}_{ab,n}^{\mu}(0)
 ({ {\mu^{\prime}}^3 \over 6\pi^2})
{1 \over v.k
+\Delta E_{ab,n}(\mu)
+i\epsilon}\,,
\eqno (4.3)
$$
$$\Delta E_{ab,n}(\mu):=
 E_{ab,n}(\mu)-
 E_{ab,n}\,.
$$
Because we only include momenta larger than $\mu$ in the gluon
propagator
giving rise to the Coulomb potential, both the wave function and the
bound state energy depend on $\mu$.

More precisely the Coulomb potential is modified by
$$
 \Delta V (\vec p , {\vec p}^{\prime} )
=
C_{f}g^{2}\theta
(\mu- \vert \vec p - {\vec p}^{\prime}\vert) {1\over{(\vec p -{\vec p}
^{\prime})^{2}}}. \eqno (4.4) $$
  Since $\mu a_{ab,n} << 1$, $\Delta V$ can be treated as a
perturbation. We may evaluate
$\Psi_{ab,n}^{\mu}$ and
$\Delta E_{ab,n}(\mu)$ at any desired order in quantum mechanical
perturbation theory.

\bigskip
In the HQET we have
$$G_{\Gamma\Gamma^{\prime}}(k):=
\int d^4x e^{-ikx}\langle
0\vert T\left\{ \bar
h^{a}_{-}{\Gamma}{h}^{b}_{+}(0)\bar {h}^{b}_{+} \Gamma^{\prime}
{h}^{a}_{-}(x)\right\} \vert 0 \rangle\,.
\eqno (4.5)
$$
At one loop

$$
  G_{\Gamma\Gamma^{\prime}}(k)=
  {G_{\Gamma\Gamma^{\prime}}}^0(k)+
  {G_{\Gamma\Gamma^{\prime}}}^{11}(k)+
  {G_{\Gamma\Gamma^{\prime}}}^{12}(k)+
  {G_{\Gamma\Gamma^{\prime}}}^{13}(k)\,. \eqno(4.6)$$
where ${G_{\Gamma\Gamma^{\prime}}}^0(k)$ stands for the tree level
contribution. It is easy to see that (4.3) matches
  ${G_{\Gamma\Gamma^{\prime}}}^0(k)$
 with the following
Wilson coefficient $C_{\Gamma}(\mu)$ and modification of the HQET
lagrangian $\Delta {\it L}$
 $$  C_{\Gamma}(\mu)=
\tilde \Psi_{ab,n}^{\ast\,\,\mu}(0)\Psi_{ab,n}^{\mu}(0)
$$
$$ \Delta {\it L}=
 {1\over 2}\Delta E_{ab,n}(\mu)\bar h^{a} h^{a}
 +{1\over 2}\Delta E_{ab,n}(\mu)\bar h^{b} h^{b}\,.
                      \eqno (4.7)
 $$

That is
$$G_{\Gamma\Gamma^{\prime}}(p)=C_{\Gamma}(\mu)G_{\Gamma\Gamma^{\prime}}(k)
\,.  \eqno(4.8)$$

\bigskip

With the modifications above the HQET lagrangian gives rise to the
following one loop contributions (see Fig. 3)

 $$
\eqalign{&
-2{G_{\Gamma\Gamma^{\prime}}}^{11}(k)=
 -2{G_{\Gamma\Gamma^{\prime}}}^{12}(k)=
  {G_{\Gamma\Gamma^{\prime}}}^{13}(k)=\cr &
-i
N_{c}tr(p_{-} \Gamma p_{+} \Gamma^{\prime})
 ({ {\mu^{\prime}}^3 \over 6\pi^2})
({1 \over v.k
+\Delta E_{ab,n}(\mu)
+i\epsilon})^2 {C_{f}\alpha\over \pi}2\mu\,.}
\eqno (4.9)
$$

\bigskip

The contributions of the full theory (4.2)
 at one loop can be written as (see Fig. 4)
$$\eqalignno{
  G_{\Gamma\Gamma^{\prime}}^{on}(p)=&
  {G_{\Gamma\Gamma^{\prime}}^{on}}^0(p)+
  {G_{\Gamma\Gamma^{\prime}}^{on}}^{11}(p)+
  {G_{\Gamma\Gamma^{\prime}}^{on}}^{12}(p)+
  {G_{\Gamma\Gamma^{\prime}}^{on}}^{13}(p) &(4.10) \cr
  {G_{\Gamma\Gamma^{\prime}}^{on}}^{11}(p):=&\int d^4x
e^{ipx} \langle 0\vert T\bigl\{ {\bar
Q^{a}} \Gamma Q^{b}(0)
\int d^4z d^4z^{\prime} {i^2\over 2}
& \cr & \times
(\bar Q^{a}\gamma_{\mu}B^{\mu}  Q^{a})(z)
(\bar Q^{a}\gamma_{\nu}B^{\nu}  Q^{a})(z^{\prime})
{\bar Q}^{b}\Gamma^{\prime}
Q^{a}(x)
 \bigr\} \vert 0 \rangle \,. &(4.11) \cr
  {G_{\Gamma\Gamma^{\prime}}^{on}}^{12}(p):=&\int d^4x
e^{ipx} \langle 0\vert T\bigl\{ {\bar
Q^{a}} \Gamma Q^{b}(0)
\int d^4z d^4z^{\prime} {i^2\over2}
 &\cr & \times
(\bar Q^{b}\gamma_{\mu}B^{\mu}  Q^{b})(z)
(\bar Q^{b}\gamma_{\nu}B^{\nu}  Q^{b})(z^{\prime})
{\bar Q}^{b }\Gamma^{\prime}
Q^{a}(x)
 \bigr\} \vert 0 \rangle \,. & (4.12) \cr
  {G_{\Gamma\Gamma^{\prime}}^{on}}^{13}(p):=&\int d^4x
e^{ipx} \langle 0\vert T\bigl\{ {\bar
Q^{a}} \Gamma Q^{b}(0)
\int d^4z d^4z^{\prime} i^2
 &\cr & \times
(\bar Q^{b}\gamma_{\mu}B^{\mu}  Q^{b})(z)
(\bar Q^{a}\gamma_{\nu}B^{\nu}  Q^{a})(z^{\prime})
{\bar Q}^{b }\Gamma^{\prime}
Q^{a}(x)
 \bigr\} \vert 0 \rangle \,. & (4.13) \cr}
$$
In order to account for momenta larger than $\mu$ we insert resolutions
of the identity between the currents,
 which we approximate by CTBS, as we did in
(3.10). Recall that the gluon propagator must be cut-off at $\mu<<
1/a_{ab,n}$ and
that almost on-shell contribution is implemented by cutting off the
relative momenta larger than $\mu^{\prime}<< 1/a_{ab,n}$.
We
obtain
 $$
\eqalign{&
-2{G_{\Gamma\Gamma^{\prime}}^{on}}^{11}(p)=
 -2{G_{\Gamma\Gamma^{\prime}}^{on}}^{12}(p)=
  {G_{\Gamma\Gamma^{\prime}}^{on}}^{13}(p)=\cr &
-i
N_{c}tr(p_{-} \Gamma p_{+} \Gamma^{\prime})
\tilde \Psi_{ab,n}^{\ast\,\,\mu}(0)
\Psi_{ab,n}^{\mu}(0)
 ({ {\mu^{\prime}}^3 \over 6\pi^2})
({1 \over v.k
+\Delta E_{ab,n}(\mu)
+i\epsilon})^2 {C_{f}\alpha\over \pi}2\mu\,.}
\eqno (4.14)
$$

\bigskip

It is remarkable that each of the contributions above
matches the corresponding one loop contribution in (4.9) with the
same modification
in the HQET lagrangian and the same Wilson coefficient (4.7). That is
(4.8) holds at one loop as well.

\bigskip

Although the matching stated as above makes perfect mathematical sense
as it stands,
it still requires some adaptation to the precise physical situation we
are concerned with. Namely, in a CTBS we wish to
separate large and small gluon momentum contributions. A CTBS is
obtained through an infinite sum of one gluon exchange diagrams only.
Consequently in the perturbative evaluation of
 $G_{\Gamma\Gamma^{\prime}}(p)$
 we
should only keep diagrams which correspond to one gluon exchange.
This means, for instance, that at one loop (4.11) and (4.12) must be
dropped, since these contributions correspond to quark and
antiquark selfenergies. Once this is done, it is easy to see that
 $G_{\Gamma\Gamma^{\prime}}(p)$
becomes $\mu$ independent if we calculate
$\Psi_{ab,n}^{\mu}$ and
$\Delta E_{ab,n}(\mu)$ at the same order in quantum
mechanical perturbation theory as we have calculated the small
momentum contributions in perturbation theory.

\bigskip

For instance,
since we used zeroth order in perturbation theory in (4.3) we have to
calculate
$\Psi_{ab,n}^{\mu}$  and
$\Delta E_{ab,n}(\mu)$ at zeroth order in quantum mechanical
perturbation theory. That is
$$
\Psi_{ab,n}^{\mu}=
\Psi_{ab,n} \quad\quad ,\quad\quad
\Delta E_{ab,n}(\mu)=0\,, \eqno (4.15)
$$
which substituted in (4.8) gives
the result
 found in [9].
At one loop order we must
 calculate
$\Psi_{ab,n}^{\mu}$ and
$\Delta E_{ab,n}(\mu)$
 at first
order in quantum mechanical perturbation theory. We obtain
$$
\Psi_{ab,n}^{\mu}=
\Psi_{ab,n} \quad\quad ,\quad\quad
\Delta E_{ab,n}(\mu)= -{C_{f}\alpha\over\pi}2\mu\,. \eqno (4.16)
$$
If we next consistenly expand
 $ {G_{\Gamma\Gamma^{\prime}}^{on}}^0(p)$
in $\alpha \mu$ at first order we see that the second
term in the expansion exactly cancels
 $ {G_{\Gamma\Gamma^{\prime}}^{on}}^{13}(p)$.
Since we dropped (4.11) and (4.12)  we must also drop their
counterparts
 in the HQET in order to keep (4.8) true.
Fortunately,
this can
be safely implemented by the addition of the following mass counterterms
in the HQET lagrangian
\footnote{$^{\dagger}$}{(4.17)
has a clear physical meaning. It arises because when integrating large
momenta
we do it from an infrared cut-off
$\mu$ onwards. After doing so, the pole mass depends on $\mu$.
However, we do not actually
work with a $\mu$ dependent mass but with the pole mass at $\mu=0$.
The $\mu$-dependent pole mass and the pole mass we work with differ by
an
amount $\delta m \sim \mu$. This difference is the origin
of (4.17). Notice that
this is much in
the same way as we use $E_{ab,n}$ instead of
$E_{ab,n}(\mu)$ and hence the effective mass term (4.7) arises.}
$$ \Delta{\it L}^{\prime}=
 -{C_{f}\alpha\over \pi}\mu \bar h^{a} h^{a} -
  {C_{f}\alpha\over \pi}\mu \bar h^{b} h^{b} \,.   \eqno (4.17)
$$

\bigskip

It is again remarkable that the mass counterterms above exactly cancel
the effective masses
$-\Delta E_{ab,n}(\mu)/2$ in  (4.7)
at this order. As the total outcome we have that at one loop the Wilson
coefficient is the same
as at zeroth order and that the HQET lagrangian must not be modified
either.
It is not clear to us whether the fact that neither the Wilson
coefficient nor the HQET lagrangian receive corrections at one loop
remains true at higher orders in perturbation theory.

\bigskip

The maching we have carried out
is exact when $m_{Q}\rightarrow \infty$. The philosophy is
analogous to the matching
carried out in heavy-light systems. However, the matching here is
done in perturbation theory for small momentum gluons only. Large
momentum
gluons are summed up in the ladder diagrams in order to account
for the CTBS.
 Then the Wilson coefficients
have to do with the wave functions instead of with anomalous
dimensions.

\bigskip

Once we have established in the proper framework above the matching
between HQET and the
CTBS, we can safely use the symmetries of the
former
in order to get information in a region where perturbation theory is not
applicable, as it was initially done in [9].

\bigskip\bigskip

{\bf 5. Conclusions}

\bigskip\bigskip

We have pointet out that there are no infinite imaginary
anomalous dimensions in the HQET.
 When the quark and the antiquark fields have
the same velocity the HQET must not be matched to perturbative QCD but
to CTBS contributions. We have shown how to do it at
one loop level.

 \bigskip \bigskip
{\bf Acknowledgements}
\bigskip
We thank E. Bagan, J. L. Goity and P. Gosdzinsky for a critical
reading of the manuscript.
This work has been supported in part by CICYT grant AEN93-0695 and
Comissionat per Universitats i Recerca de la Generalitat de Catalunya.
A.P. acknowledges a fellowship from CIRIT. \bigskip

\bigskip
\centerline{\bf References}
\bigskip

\item{[1]} M.B. Voloshin and M.A. Shifman, {\it Yad. Fiz.} {\bf 45}
(1987) 463 [Sov. J. {\it Nucl. Phys.} {\bf 45} (1987) 292].
 \item{} H.D.
Politzer and M.B. Wise, {\it Phys. Lett.} {\bf B206}
(1988) 681;  {\it Phys. Lett.} {\bf B208} (1988) 504.
\item{} N. Isgur and M.B. Wise, {\it Phys. Lett.} {\bf B232}
(1989) 113;  {\it Phys. Lett.} {\bf B237} (1990) 527.
\item{} E. Eichen and B. Hill, {\it Phys. Lett.} {\bf B234} (1990)
511.
\item{} H. Georgi,
{\it Phys. Lett.} {\bf B240} (1990) 447.
\item{} B. Grinstein, {\it Nucl. Phys.} {\bf B339} (1990) 253.

\item{[2]} B. Grinstein, in {\it Proceedings of the Workshop on High
Energy Phenomenology}, Mexico City Mexico, Jul. 1-12, 1991, 161-216
and in {\it Proceedings Intersections between particle and nuclear physics}
Tucson 1991, 112-126.
\item{} T. Mannel, {\it Chinese Journal of Physics} {\bf 31} (1993) 1.
\item{} M. Neubert, {\it Phys. Rep.} {\bf245} (1994) 259.

\item{[3]} M. B. Voloshin,
{\it Nucl. Phys.} {\bf B154} (1979) 365.

\item{[4]} H. Leutwyler,
{\it Phys. Lett.} {\bf B98} (1981) 447.

\item{[5]} S. Titard and F.J. Yndur\'ain, in press in {\it Phys.Rev.}
FTUAM 94-34.

\item{[6]} I.I. Bigi, M.A. Shifman, N.G. Uraltsev and A.I. Vainshtein,
{\it Phys. Rev.} {\bf D50} (1994) 2234.

\item{[7]}
 W. Kilian, T. Mannel and T. Ohl,
{\it Phys. Lett.} {\bf B304} (1993) 311

\item{[8]} J. Soto and R. Tzani, {\it Int. J. Mod. Phys.} {\bf A9}
(1994) 4949.

\item{[9]} A. Pineda and J. Soto, {\it 'Heavy Quark Hadronic Lagrangian
for S-Wave Quarkonia'}, UB-ECM-PF-94/19, hep-ph 9409216.

\item{[10]} B. Grinstein, W. Kilian, T. Mannel and M. Wise, {\it Nucl.
Phys.} {\bf B363} (1991) 19.
\item{} W. Kilian, P. Manakos and T.
Mannel, {\it Phys. Rev.} {\bf D48} (1993) 1321.

\item{[11]} W.E. Caswell and G.P. Lepage, {\it Phys. Lett.} {\bf 167B}
(1986) 437; G.P. Lepage and B.A. Thacker, {\it Nucl. Phys.(Proc.
Suppl.)} {\bf 4} (1988) 199.

\item{[12]} C.T.H. Davies, K. Hornbostel, A. Langau, G.P. Lepage, A.
Liedsey, C.J. Morningstar, J. Shigemitsu and J. Sloan, {\it Phys. Rev.
Lett.} {\bf 73} (1994) 2654;
C.T.H. Davies, K. Hornbostel, A. Langau, G.P. Lepage, A.
Liedsey, J. Shigemitsu and J. Sloan, {\it Phys. Rev.}
{\bf D50} (1994) 6963.

\item{[13]} E. Bagan and P. Gosdzinsky, {\it Phys. Lett.} {\bf B320}
(1994) 123.

\vfill
\eject

\magnification=1200

Caption 1. Diagrams contributing to $G_{\Gamma}$ at one loop in the
HQET theory. The heavy
quark field carries residual momentum $k_1$ and the heavy antiquark
field residual momentum $k_2$.
\bigskip\bigskip

Caption 2. Diagrams contributing to $G_{\Gamma}$ at one loop in the full
theory.
The quark field carries momentum $p_1$ and the antiquark
field momentum $p_2$. The double line means almost on-shell, the wavy
line means soft gluon and the square means summing up all the ladder
diagrams.
\bigskip\bigskip

Caption 3. Diagrams contributing to $G_{\Gamma \Gamma^{\prime}}$ at one
loop in the HQET theory. The heavy quark anti-quark current carries
 momentum $k$.
\bigskip\bigskip

Caption 4. Diagrams contributing to $G_{\Gamma \Gamma^{\prime}}^{on}$ at
one loop in the full theory. The current carries momentum $p$.
\bigskip\bigskip

\end